\documentclass[a4paper,11pt]{article}

\usepackage{jinstpub} 
\usepackage{lineno}
\usepackage[skip=0pt]{subcaption}

\newcommand{\nuebar}{\ensuremath{\overline{\nu}_{e}} }
\newcommand{\uFive}{$^{235}$U}

\newcommand{\Li}{$^{6}$Li}

\title{Calibration strategy of the PROSPECT-II detector with external and intrinsic sources}

\author[g]{M.~Andriamirado,}
\author[r]{A.~B.~Balantekin,}
\author[i]{C.~D.~Bass,}
\author[j]{D.~E.~Bergeron,}
\author[h]{E.~P.~Bernard,}
\author[h]{N.~S.~Bowden,}
\author[k]{C.~D.~Bryan,}
\author[p]{R.~Carr,}
\author[h]{T.~Classen,}
\author[k]{A.~J.~Conant,}
\author[l,o]{A.~Delgado,}
\author[c]{M.~V.~Diwan,}
\author[d]{M.~J.~Dolinski,}
\author[e]{A.~Erickson,}
\author[s]{B.~T.~Foust,}
\author[s]{J.~K.~Gaison,}
\author[l,o]{A.~Galindo-Uribarri,}
\author[l,o]{C.~E.~Gilbert,}
\author[c]{S.~Gokhale,}
\author[a]{C.~Grant,}
\author[b,c]{S.~Hans,}
\author[m]{A.~B.~Hansell,}
\author[s]{K.~M.~Heeger,}
\author[l,o]{B.~Heffron,}
\author[c]{D.~E.~Jaffe,}
\author[d]{S.~Jayakumar,}
\author[c]{X.~Ji,}
\author[n]{D.~C.~Jones,}
\author[f]{J.~Koblanski,}
\author[a]{P.~Kunkle,}
\author[d]{C.~E.~Lane,}
\author[s]{T.~J.~Langford,}
\author[j]{J.~LaRosa,}
\author[g]{B.~R.~Littlejohn,}
\author[l,o]{X.~Lu,}
\author[f]{J.~Maricic,}
\author[h]{M.~P.~Mendenhall,}
\author[f]{A.~M.~Meyer,}
\author[f]{R.~Milincic,}
\author[l]{P.~E.~Mueller,}
\author[j]{H.~P.~Mumm,}
\author[n]{J.~Napolitano,}
\author[d]{R.~Neilson,}
\author[s]{J.~A.~Nikkel,}
\author[j]{S.~Nour,}
\author[g]{J.~L.~Palomino,}
\author[q]{D.~A.~Pushin,}
\author[c]{X.~Qian,}
\author[h]{C.~Roca,}
\author[c]{R.~Rosero,}
\author[k]{M.~Searles,}
\author[s]{P.~T.~Surukuchi,}
\author[h]{F.~Sutanto,}
\author[j]{M.~A.~Tyra,}
\author[l,o]{D.~Venegas-Vargas,}
\author[d]{P.~B.~Weatherly,}
\author[s]{J.~Wilhelmi,}
\author[q]{A.~Woolverton,}
\author[c]{M.~Yeh,}
\author[c]{C.~Zhang}
\author[h]{and X.~Zhang}

\affiliation[a]{Department of Physics, Boston University, Boston, MA, USA}
\affiliation[b]{Department of Chemistry and Chemical Technology, Bronx Community College, Bronx, NY, USA}
\affiliation[c]{Brookhaven National Laboratory, Upton, NY, USA}
\affiliation[d]{Department of Physics, Drexel University, Philadelphia, PA, USA}
\affiliation[e]{George W.\,Woodruff School of Mechanical Engineering, Georgia Institute of Technology, Atlanta, GA, USA}
\affiliation[f]{Department of Physics and Astronomy, University of Hawaii, Honolulu, HI, USA}
\affiliation[g]{Department of Physics, Illinois Institute of Technology, Chicago, IL, US}
\affiliation[h]{Nuclear and Chemical Sciences Division, Lawrence Livermore National Laboratory, Livermore, CA, USA}
\affiliation[i]{Department of Physics, Le Moyne College, Syracuse, NY, USA}
\affiliation[j]{National Institute of Standards and Technology, Gaithersburg, MD, USA}
\affiliation[k]{High Flux Isotope Reactor, Oak Ridge National Laboratory, Oak Ridge, TN, USA}
\affiliation[l]{Physics Division, Oak Ridge National Laboratory, Oak Ridge, TN, USA}
\affiliation[m]{Department of Physics, Susquehanna University, Selinsgrove, PA, USA}
\affiliation[n]{Department of Physics, Temple University, Philadelphia, PA, USA}
\affiliation[o]{Department of Physics and Astronomy, University of Tennessee, Knoxville, TN, USA}
\affiliation[p]{Department of Physics, United States Naval Academy, Annapolis, MD, USA}
\affiliation[q]{Institute for Quantum Computing and Department of Physics, University of Waterloo, Waterloo, ON, Canada}
\affiliation[r]{Department of Physics, University of Wisconsin, Madison, WI, USA}
\affiliation[s]{Wright Laboratory, Department of Physics, Yale University, New Haven, CT, USA}

\emailAdd{prospect.collaboration@gmail.com}

\abstract{This paper presents an energy calibration scheme for an upgraded reactor antineutrino detector for the Precision Reactor Oscillation and Spectrum Experiment (PROSPECT). The PROSPECT collaboration is preparing an upgraded detector, PROSPECT-II (P-II), to advance capabilities for the investigation of fundamental neutrino physics, fission processes and associated reactor neutrino flux, and nuclear security applications. P-II will expand the statistical power of the original PROSPECT (P-I) dataset by at least an order of magnitude. The new design builds upon 
previous P-I design and focuses on improving the detector robustness and long-term stability to enable multi-year operation at one or more sites. The new design optimizes the fiducial volume by elimination of dead space previously occupied by internal calibration channels, which in turn necessitates the external deployment. In this paper, we describe a calibration strategy for P-II. The expected performance of externally deployed calibration sources is evaluated using P-I data and a well-benchmarked simulation package by varying detector segmentation configurations in the analysis. The proposed external calibration scheme delivers a compatible energy scale model and achieves comparable performance with the inclusion of an additional AmBe neutron source, in comparison to the previous internal arrangement. Most importantly, the estimated uncertainty contribution from the external energy scale calibration model meets the precision requirements of the P-II experiment.
}

\begin{document}
\maketitle
\flushbottom

\section{Introduction}\label{sec:intro}
The Standard Model (SM) of physics has been remarkably successful in explaining fundamental particles and forces of nature, yet it has been established that the model is incomplete. There are several indications of physics beyond the SM in the neutrino sector. The discovery of neutrino oscillations~\cite{SNO,SuperK} demonstrated that neutrinos have masses, a sign of new physics since neutrinos are massless in the SM. In addition, there are several experimental hints of the existence of sterile neutrinos~\cite{Gallex,sage,LSND,miniboone,BEST} that have motivated searches for sterile neutrinos at very short distance ($\sim$10~m) from nuclear reactor cores, such as the short baseline neutrino experiment conducted by the PROSPECT Collaboration~\cite{prospect}. 

PROSPECT is the first short-baseline reactor neutrino physics experiment in the United States since the Savannah River experiments of Reines {\it et al.}~\cite{Reines}. This detection instrument was deployed at the 85~MW High Flux Isotope Reactor (HFIR) at Oak Ridge National Laboratory. PROSPECT was designed to address two most prominent anomalous results facing reactor neutrino physics: the deficit of observed neutrino flux relative to prediction~\cite{fluxAnomaly} and a spectral discrepancy with respect to theoretical models around the 5-7~MeV range~\cite{RENO,DC,DYB}. PROSPECT experiment took advantage of the enormous flux of MeV-scale pure electron-type antineutrinos predominantly produced via beta decay of fission daughters of \uFive ~from the highly-enriched uranium (HEU) research reactor core at HFIR. The detector was centered at a baseline of 7.9 m from the reactor core to perform sterile neutrino searches with a mass splitting on the order of 1 eV$^2$. The experimental setup is well optimized for conducting a high precision measurement of the reactor $^{235}$U antineutrino spectrum. The collaboration has reported a high statistics spectrum measurement of reactor $^{235}$U antineutrinos~\cite{spec}, ruled out a substantial region of the allowed sterile parameter space with high level of confidence~\cite{osc}, and performed joint spectral analyses with other reactor experiments~\cite{DBYjoint,STEjoint}. 

\begin{figure}[ht]
\centering
    \includegraphics[width=0.9\textwidth]{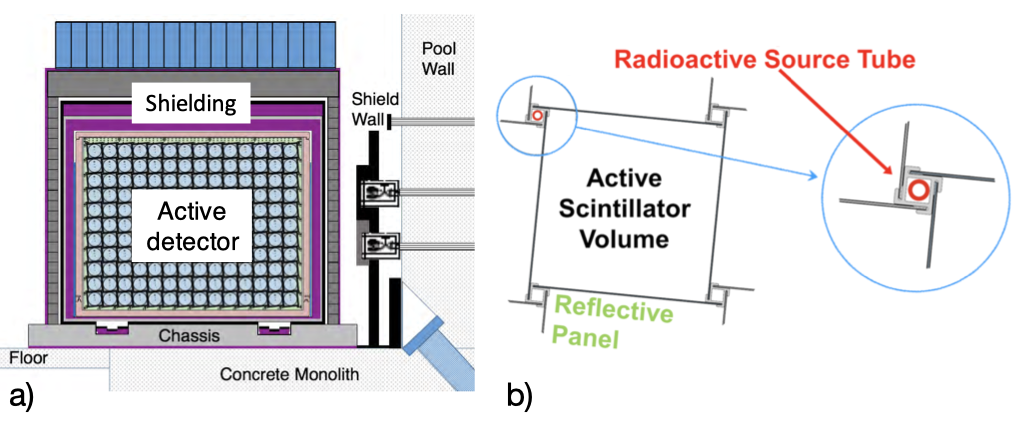}
    \caption{a) side view of the P-I detector active volume divided into a 14$\times$11 optical grid. b) cartoon of the segment layout, creating space to accommodate source tubes.  }
    \label{fig:detector}
\end{figure}

The P-I detector design contains $\sim$4 tons of \Li-doped liquid scintillator~\cite{liquid}, optically separated by reflective panels~\cite{lowmass} into a 14$\times$11 array with 1176~mm $\times$ 145~mm $\times$ 145~mm segment readout by photomultiplier tubes (PMTs) on each end, that covers baselines approximately between 7 and 9 meters. Each optical segment is tilted by 5.5$^\circ$ and positioned in an offset array to accommodate source tubes in the space between adjacent segments. Figure~\ref{fig:detector} shows the side view of the segmented detector and a tilted segmentation design that accommodates calibration source tubes.
Inside the source tubes, calibration source capsules are attached to a belt to be driven by external motors to given positions along the length of the segment during calibration, and retracted to a home position outside the detector during data taking~\cite{cal_paper}.

As an on-surface reactor antineutrino detector that is subject to high rates of reactor-related and cosmogenic backgrounds, efficient selection of neutrino signal requires particle differentiation of time-correlated positron and neutron events produced by the inverse beta decay (IBD) process:
\begin{equation}
\nuebar + p \to e^+ + n.
\label{eqn:IBD}
\end{equation}  

For this purpose, efficient pulse shape discrimination (PSD) techniques are required. The digitized waveforms of the scintillation light, produced via the interactions of ionizing particles in the liquid scintillator, are characterized by different decay components. The decay time depends on the atomic and molecular structure of radiative states. Signals generated by interactions of lower-ionization-density events, such as $\beta$-particles, will be very different from those produced by higher-ionization-density ones like proton recoil events. In our current analysis, a simple PSD parameter is defined by the ratio of the area contained in the tail of the pulse over the whole area, ensuring adequate separation between electron-like and proton-like interactions~\cite{LS}. Positron signals from the IBD are thus distinguishable from thermal neutron captures on $\mathrm{^{6}Li}$ (which generates a triton and an $\alpha$-particle). This particle discriminating ability, combined with spatial and temporal correlations of the positron and neutron, offers a powerful handle for IBD interaction identification and background suppression. 

Positrons in the IBD process carry most of the kinetic energy of the neutrinos. The PROSPECT experiment, as with other IBD-based neutrino experiments, measures energy depositions from the positron and pair of annihilation photons, referred to as the visible energy. The spectral discrepancy of the measured visible energy spectrum from the theoretical prediction reported in reactor neutrino experiments, such as RENO~\cite{RENO}, Double Chooz~\cite{DC}, Daya Bay~\cite{DYB}, is dependent on a well-characterized energy response, thus requiring a precision energy calibration of the detector. 

The P-I calibration campaign comprised periodic deployment of internal $\gamma$-ray and neutron sources during the detector data-taking period. The internal radioactive source data, together with cosmogenic ${}^{12}$B decays, were analyzed to determine the detector response model, the PSD particle identification, the overall energy scale, and nonlinearity~\cite{prd}. With the detector precisely calibrated, the energy, position, and particle type of the interaction can be accurately reconstructed.

PROSPECT was successful despite early termination of operation due to a combination of PMT divider failure and unexpected HFIR downtime. An upgraded P-II detector design has multiple improvements to ensure the long-term stability and robustness of the detector, and to facilitate its possible relocation to an low-enriched-uranium (LEU) reactor after an initial deployment at HFIR. With the goal of collecting a $^{235}$U reactor antineutrino dataset with statistical uncertainties lower than experimental systematics, P-II offers a unique opportunity to achieve a higher statistics measurement of the $^{235}$U antineutrino spectrum, a precise determination of the absolute flux of antineutrinos from $^{235}$U fission, and increased sensitivity for sterile neutrino searches~\cite{p2physics}. 

\begin{figure}[h]
\centering
\begin{subfigure}[b]{\textwidth}
    \centering
   \includegraphics[width=0.9\textwidth]{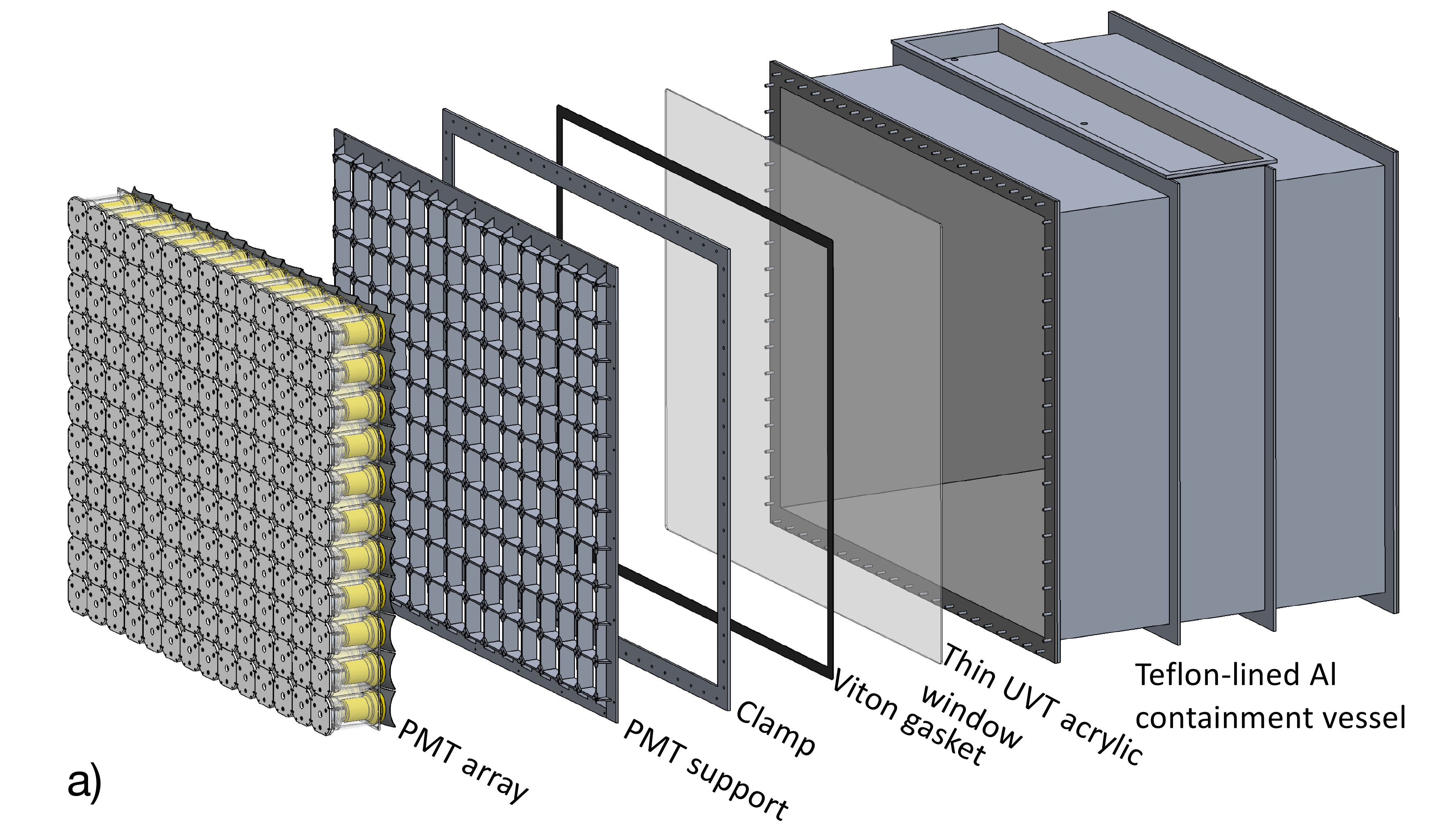}
\end{subfigure}
\begin{subfigure}[b]{\textwidth}
    \centering
   \includegraphics[width=0.9\textwidth]{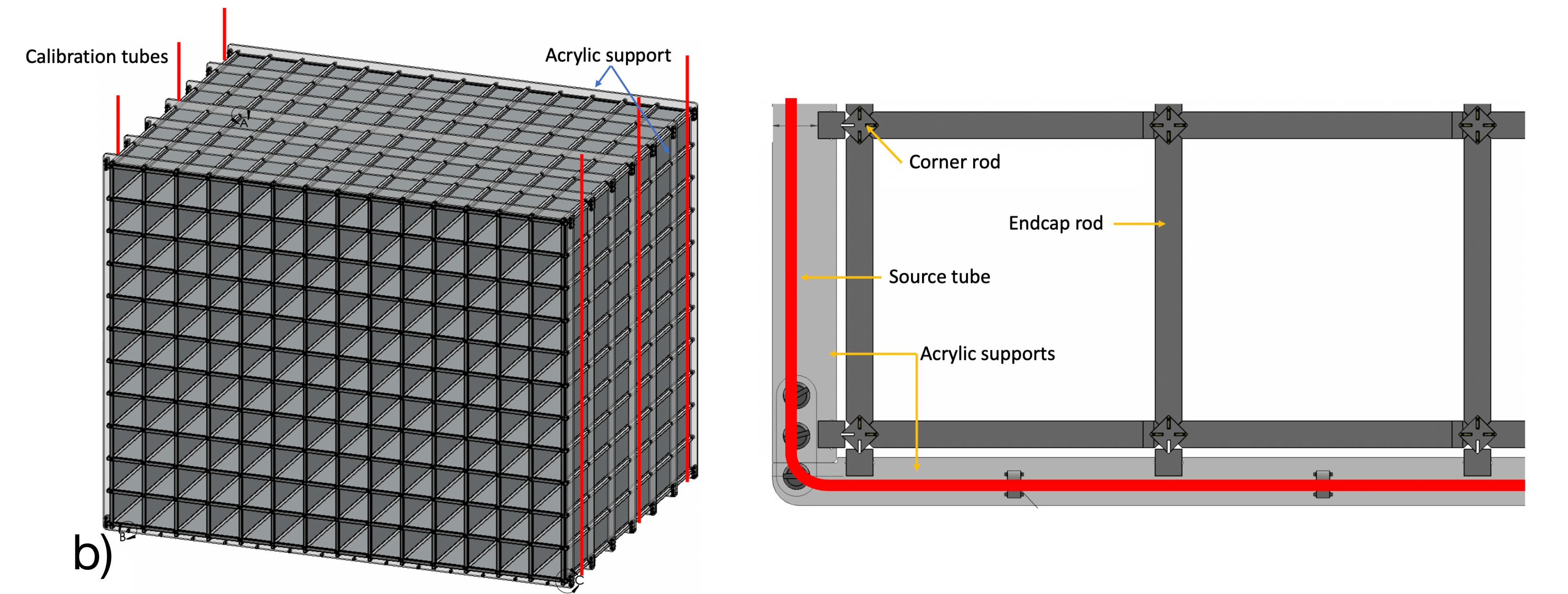}
\end{subfigure}
\caption{ a) An exploded view of a preliminary P-II detector design. PMT array is separated from the liquid scintillator containment vessel by a thin acrylic window. b) Proposed placement of external calibration source tubes along the perimeter of the segment array, as shown in red. }
\label{fig:prospect-II}
\end{figure}

One of the main upgrades in the P-II detector is the separation of PMTs from the liquid scintillator volume to eliminate the possibility of liquid scintillator leakage into the PMT housings. This change allows for a simpler, fully-sealed liquid scintillator tank structure with minimal penetrations. Calibration access to the interior of the optical lattice would increase the complexity and risks associated with the tank design. Therefore we explore a calibration scheme that uses deployable sources located outside the optical lattice but inside the tank. Figure~\ref{fig:prospect-II} shows an exploded view of a preliminary P-II detector design and a proposed calibration setup. We will refer to this scheme as external calibration throughout the rest of this paper.

Given the precision goals of P-II, it is essential to confirm that energy scale systematic uncertainties can be sufficiently constrained using an external calibration scheme. The ability to understand the response of highly segmented detectors, like P-I and P-II, requires detailed Monte Carlo (MC) simulation models and data from a range of calibration inputs across the entire relevant energy range. When P-I was designed, we chose to deploy internal calibration sources as a conservative approach to ensure a precise energy scale. Calibration analyses for P-I demonstrated that excellent agreement between data and MC could be achieved, and that a uniform energy-scale response could be established across the entire detector. This outcome enables us to explore the feasibility of a data-driven external calibration scheme for the P-II detector and its possible applications to other similar segmented detectors. 

The study presented in this paper aims to demonstrate the feasibility of an energy calibration scheme for a multi-segmented array such as P-II, eliminating the need to deploy internal radioactive sources. The goal of the energy calibration is a relative systematic uncertainty less than or comparable to the expected P-II statistical uncertainty. With recent interests in multi-segmented detectors on the scale of PROSPECT for neutrino physics or nuclear non-proliferation monitoring purposes~\cite{Bernstein:2019hix}, external source deployment provides a less intrusive yet versatile way to calibrate liquid or solid detectors. Improved and simplified calibration techniques for multi-detector arrays is a goal shared with other fields of subatomic physics such as intermediate-energy nuclear reaction studies that utilize large 4$\pi$ multi-detector arrays. Detailed investigation of the interaction of various species with the detection medium as a function of E, Z and even A~\cite{Horn} has resulted in the characterization of the light output for a variety of nuclear species with a single quenching constant considerably simplifying the calibration process. It is worth noting that in the case of PROSPECT the detection medium is common to all segments; the segments, by construction, have very similar optical properties; and the range of energies of interest is modest (100 keV to 10 MeV) and adequately covered with conventional sources. The challenge however is the use of only external low energy sources, the low-Z detection medium, and additional passive shielding, resulting in reduction on the photopeak yields and decreased event multiplicities. 

In this paper, we present in section~\ref{sec:method} a review of the calibration procedure in P-I, followed by a description of the proposed P-II calibration procedure. In section~\ref{sec:result}, we discuss the results and studies on additional concerns introduced by external calibration configuration and P-II upgrades. The results are summarized in section~\ref{sec:conclusion}.

\section{Methodology}
\label{sec:method}
\begin{figure}[ht]
\captionsetup[subfigure]{labelformat=empty}
\begin{subfigure}[b]{0.49\textwidth}
   \includegraphics[width=\textwidth]{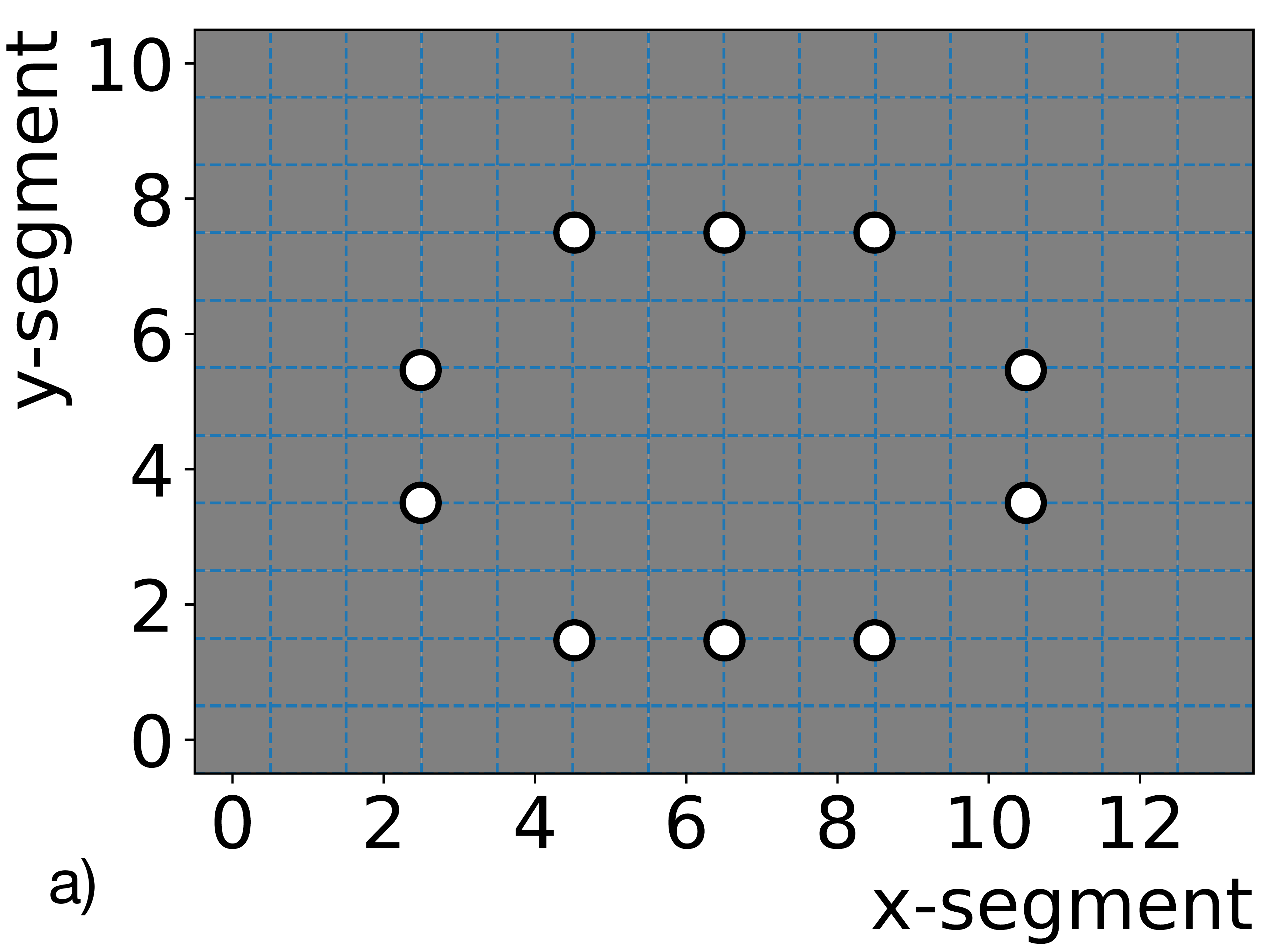}
\end{subfigure}
\hfill
\begin{subfigure}[b]{0.49\textwidth}
   \includegraphics[width=\textwidth]{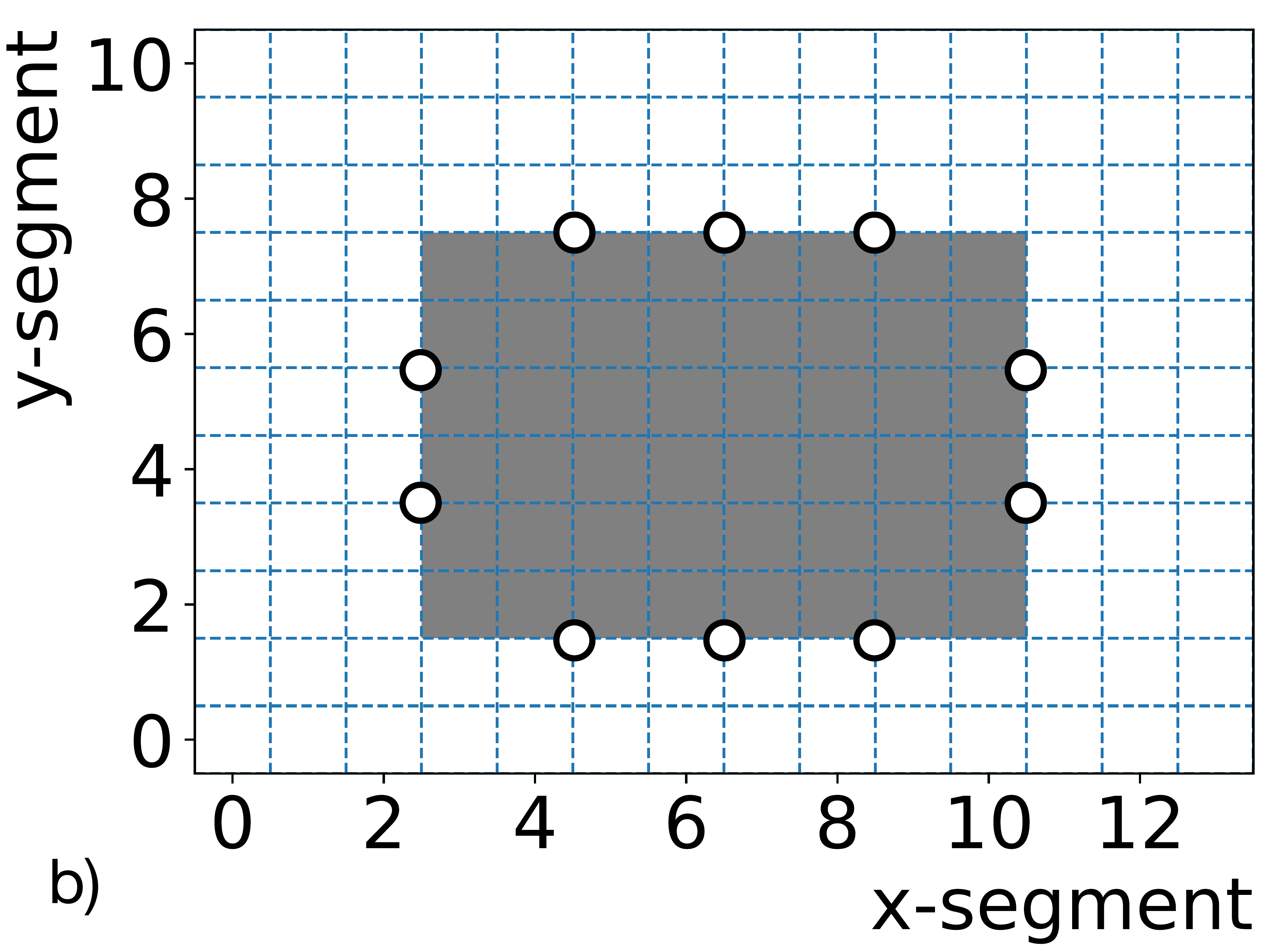}
\end{subfigure}
\caption{Illustration of the concept for using P-I internal calibration data to study "external-like" calibration response, in an ideal case where all detector segments were taking data. a) P-I internal source configuration. Active segments are represented in gray. b) P-II "external source" configuration. White segments are excluded from the calibration analysis. Source location indicated by white circles are for illustration purposes. Only one source is deployed at a time.  }
\label{fig:segs}
\end{figure}

The P-I detector design consists of 154 optically-isolated segments in a 14$\times$11 grid. This segmentation enables a data-driven study of P-II type external calibration performance using existing P-I calibration data and MC. Figure~\ref{fig:segs} shows a cross-sectional view of the detector segments and illustrates two calibration schemes: a P-I type internal calibration and a representation of the external calibration. An approximation of a P-II type external calibration is achieved by selecting data from an inner portion of the detector and excluding data from outer segments, without any effect on signal triggering and data acquisition. With such segment configuration, existing P-I calibration data and MC can be used to effectively perform external calibration analysis. Using P-I internal calibration as a benchmark, this study evaluates the performance of the calibration using externally deployed radioactive sources and intrinsic cosmogenic induced $^{12}$B decay events in P-II. 

Calibration campaigns consist of sequential deployment of a set of $\gamma$-ray and neutron sources at various locations inside the detector. A detailed detector model based on measured detector geometry and materials is built in \textsc{Geant4}~\cite{GEANT4} with tunable energy response parameters. The best-fit energy response model is determined by minimizing data-MC $\chi^2$ values. Each step is described in detail in the following subsections.

\subsection{Calibration data}\label{sec:source}
As described above, P-I calibration data can be used for an external calibration study by analyzing the response of a subset of segments, as shown in figure~\ref{fig:segs}. There were four calibration campaigns throughout the operation of the P-I detector. Each calibration campaign deployed various $\gamma$-ray and neutron sources, see table~\ref{tab:sources}. All $\gamma$-rays and neutron sources provided $\gamma$-rays of known energy used for calibration. In addition, the positron annihilation in the $^{22}$Na source allows study of annihilation $\gamma$-rays transport. Neutron production from spontaneous fission in the $^{252}$Cf source serves as a source of tagged, internally produced, ~2.22 MeV $\gamma$-rays from the p(n,$\gamma$)d reaction. The AmBe neutron source provides~4.4 MeV $\gamma$-rays via $\alpha$-particle capture on $^{9}$Be. The PROSPECT collaboration acquired the AmBe source near the end of data-taking and only used it for data-MC validation purposes in P-I analysis. Additionally, intrinsic cosmogenically produced $^{12}$B (predominantly $\beta$-decays with an energy up to 13.4~MeV) was used to establish detector response in the higher energy region.
\begin{table}[h]
\centering
\caption{Calibration sources and their $\gamma$-ray energy. The AmBe source was acquired near the end of data-taking period and was only used for data-MC validation purposes in previous P-I analyses. }
\setlength{\tabcolsep}{4pt} % default value is 6pt
\begin{tabular}{c c c c }

\hline\hline
\textbf{Source} & \textbf{Type} & \textbf{$\gamma$-ray Energy (MeV)}  & \textbf{Activity} \\ [0.5ex] 
\hline
$^{137}$Cs & $\gamma$-ray  & 0.662 & 0.1 $\mu$Ci \\  
$^{22}$Na & $\gamma$-ray & 2x 0.511, 1.275 & 0.1 $\mu$Ci \\
$^{60}$Co & $\gamma$-ray  & 1.173, 1.332 & 0.1 $\mu$Ci \\ 
$^{252}$Cf & neutron & 2.223 (n-H capture) & 866 n/s \\ 
AmBe & neutron & 4.4 & 70 n/s\\ [0.5ex]
\hline
\end{tabular}
\label{tab:sources}
\end{table}

\begin{figure}[ht]
\captionsetup[subfigure]{labelformat=empty}
\begin{subfigure}[h]{0.49\textwidth}
    \includegraphics[width=\textwidth]{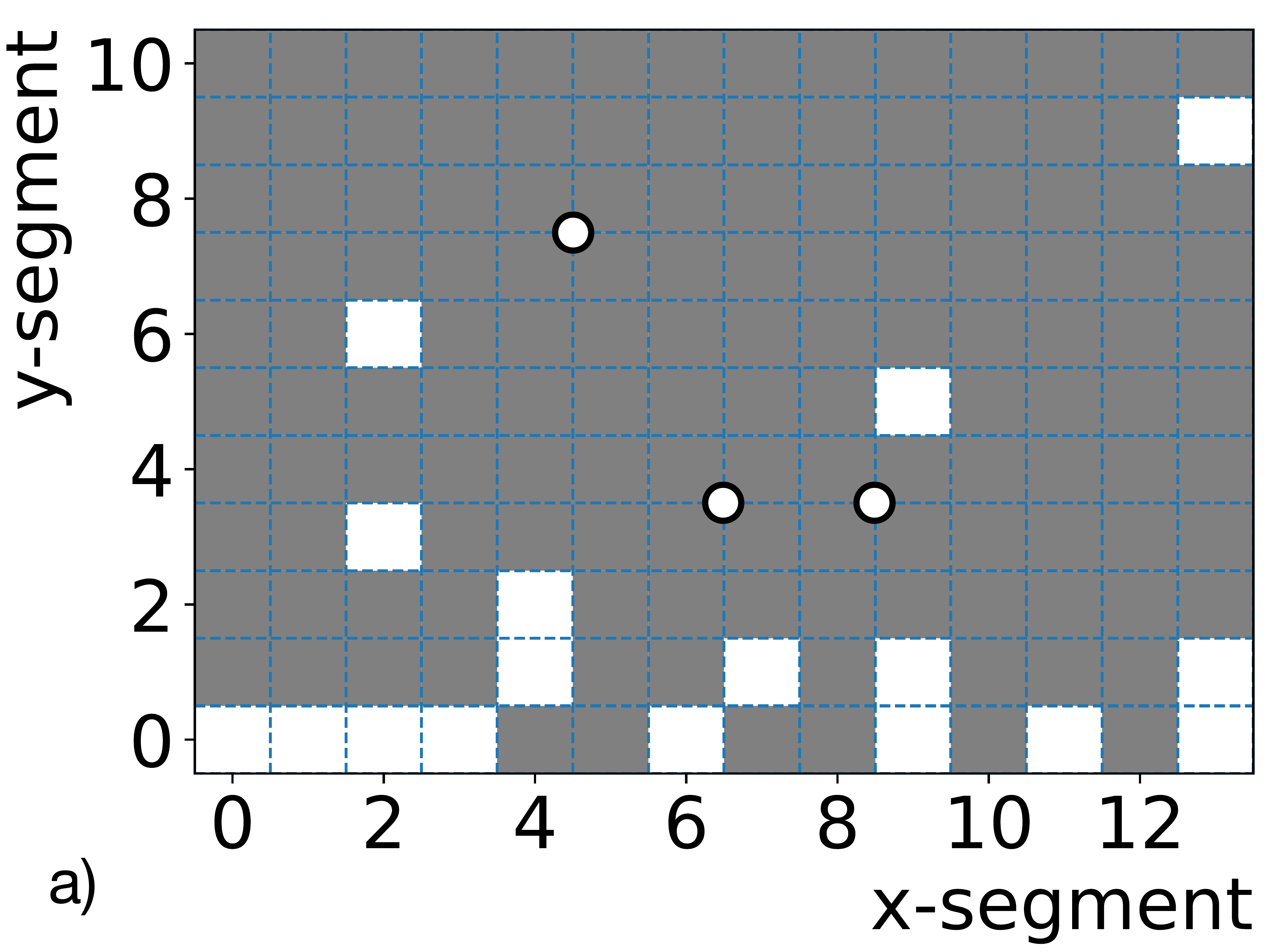}
    \caption{}
    \label{fig:segs_exp_a}
\end{subfigure}
\hfill
\begin{subfigure}[h]{0.49\textwidth}
    \includegraphics[width=\textwidth]{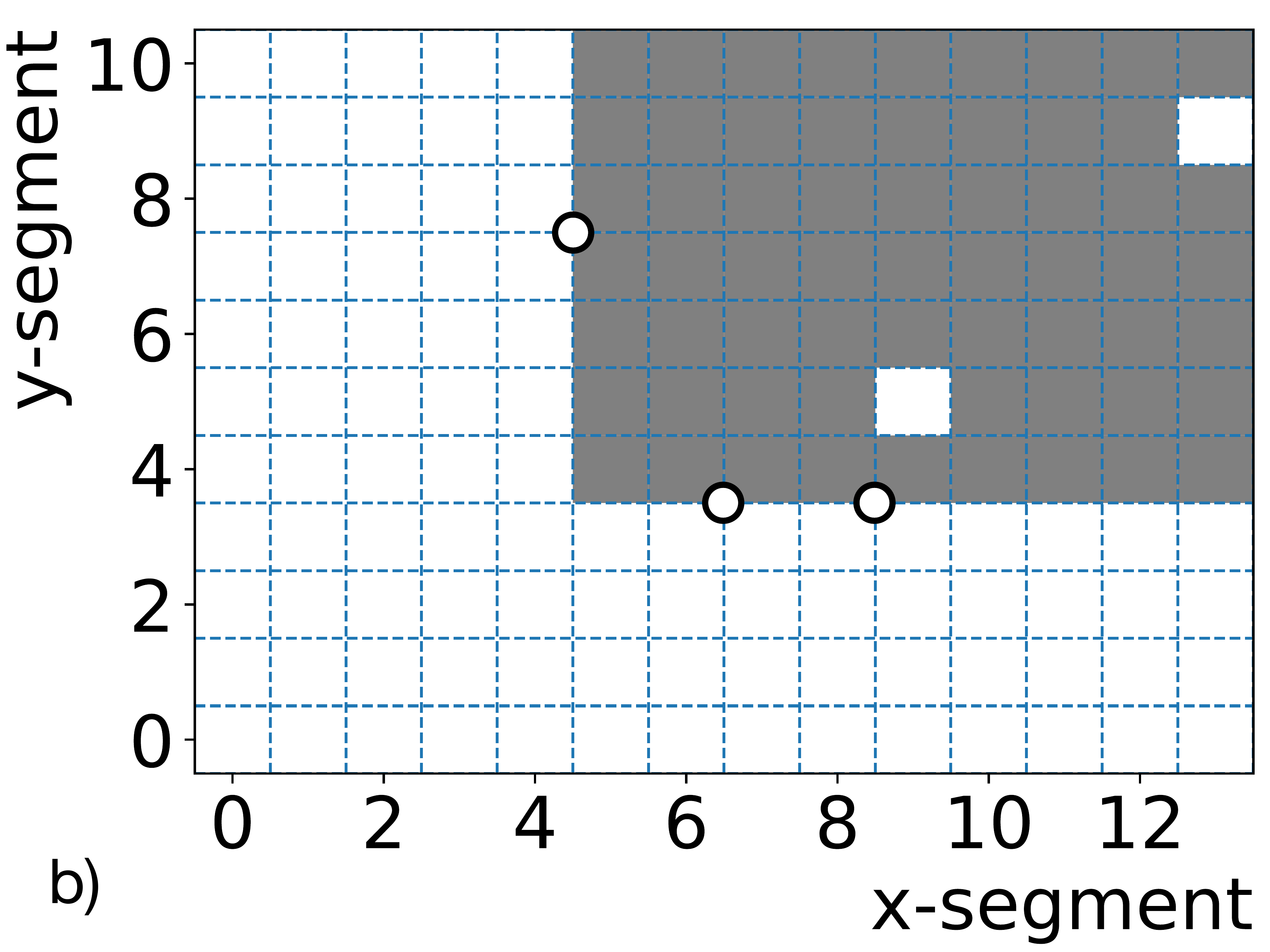}
    \caption{}
    \label{fig:segs_exp_b}
\end{subfigure}
\caption{ Illustration of the actual use of P-I internal calibration data to study "external-like" calibration response. This differs from the idealized case of figure~\ref{fig:segs} because some detector segments were shut off at the time of calibration.  White circles indicate available source locations. a) Actual P-I radioactive source and segment layout.  b) Configuration used in this work to mimic external calibration. }
\label{fig:segs_exp}
\end{figure}

During P-I calibration campaigns, the sources were deployed at various source tube locations inside the detector at seven positions 20~cm apart along the length of the segments. The center source deployment typically ran for 10 minutes mainly for energy scale analysis, while off-center deployments were usually 5 minutes for edge effect studies. The AmBe source was deployed more than 10 hours overnight to accumulate sufficient event statistics. The layout of segments and source locations evaluated here are shown in figure~\ref{fig:segs_exp}. Segments represented with white squares in figure~\ref{fig:segs_exp_a} were non-operational at the time of calibration due to PMT voltage divider instabilities. Data from many operational segments are excluded from the calibration analysis depicted in figure~\ref{fig:segs_exp_b} to mimic the proposed P-II external calibration scheme. Note that particle interactions in all non-operational or analysis-excluded segments are tracked in the supporting MC simulations.

Calibration events are identified via selection criteria based on PSD particle identification, the number of segments with energy depositions above a 90~keV energy threshold (event multiplicity), and spatial and temporal correlation of particles. Electron-like events from the $\gamma$-ray sources are selected within 3$\sigma$ of the mean electron-like PSD band as a function of energy~\cite{zhang_thesis}. A typical integral $\gamma$-ray rate of $\mathcal{O}$(1000) counts per second for the entire energy range is observed, after all selection cuts have been applied. $^{252}$Cf events are selected by identifying a correlated pair of a prompt $\gamma$-ray from $^{252}$Cf fission followed by a 2.22~MeV $\gamma$-ray from the p(n,$\gamma$)d thermal neutron capture interaction. The time coincidence cut requires the n-H capture $\gamma$-ray to be within 200~$\mu$s of the prompt $\gamma$-ray. $^{12}$B event tagging, via the $^{12}$C(n,p)$^{12}$B interaction, requires a correlated prompt single-hit proton-like recoil event within the energy range from 0.7~MeV$_{ee}$ to 10~MeV$_{ee}$, followed by a delayed signal that has electron-like PSD with energy less than 15 MeV and multiplicity $\leq{3}$. The time difference between the prompt and the delayed signals is required to be between 3 and 30~ms, and the distance separation is required to be less than 12~cm. 

\subsection{Parameterization of the light output and PROSPECT \textsc{Geant4} simulation}
Birks' law~\cite{Birks_1951} is an empirical description of the light yield per path length as a function of the energy loss per path length for a particle traversing a scintillator. Birks proposed a simple expression for the light output which took into account the quenching of light output for cases of very large energy loss. Although other P-I analyses employ a second order Birks’ model, it was found that the current data are well reproduced by a first order model.

A \textsc{Geant4}-based Monte Carlo simulation package was developed for the P-I detector. The MC package has been extensively benchmarked and reproduces detector response well~\cite{prd}. The MC simulates particles traversing the liquid scintillator, converting the deposited energy to scintillation light via a nonlinear relation. Comprehensive calibration data collected with neutron and $\gamma$-ray radioactive sources and intrinsic backgrounds, spanning the entire PROSPECT energy scale, has been utilized to tune MC simulation and accurately reproduce P-I scintillator energy response nonlinearities.  

In \textsc{Geant4}, the energy deposition is accumulated with each simulation step (G4Step) along particle tracks. We calculate \textit{dE/dx} in each G4Step and then apply effective nonlinear corrections to calculate the resulting quenched scintillation emission. The fractional conversion rate of true deposited energy in the scintillator to scintillation light is calculated and summed at each step~\cite{zhang_thesis}:

\begin{equation}\label{eq:convert}
E_{MC} = A\sum_i (E_{scint,i}(k_{B})+E_{c,i}(k_C)),
\end{equation}
where $E_{scint,i}(k_{B})$ is the scintillation light produced in the scintillator at each simulation step $i$, following Birks' empirical model of light yield, and $E_{c,i}(k_C)$ is light yield from the Cerenkov process, and $A$ is an overall normalization factor. Both $k_B$ and $k_C$ are effective parameters that incorporate scintillator properties as well as other detector effects that produce non-proportional response. Birks' law is parameterized as follows~\cite{Birks_1951}:
\begin{equation}\label{eq:birks}
\frac{dE_{scint}}{dx} = \frac{\frac{dE}{dx}}{1+k_{B}\frac{dE}{dx}},
\end{equation}
where $k_{B}$ is the effective first-order Birks's constant and $dE/dx$ is the true deposited energy in that step. Cerenkov light production, absorption, and subsequent scintillation photon re-emission in simulation step $i$ is modelled as 
\begin{equation}\label{eq:ceren}
E_{c} = k_{C}\int_{\lambda} \frac{2\pi\alpha z^2}{\lambda} \Big(1-\frac{1}{\beta^2n^2(\lambda)}\Big) E_\lambda d\lambda,
\end{equation}
where the integrand is the number of Cerenkov photons emitted per unit wavelength, $\alpha$ is the fine structure constant, $z$ is the particle's electric charge, $\beta$ is the speed of the particle, $n(\lambda)$ is the index of refraction of the medium, $E_\lambda$ is the energy of those Cerenkov photons, and $k_C$ is an effective normalization parameter that scales Cerenkov light production with respect to a default estimate based on constant scintillator refractive index assumptions for wavelengths in 200-700~nm range.  

We simulated $1\times{10}^6$ and $4\times{10}^5$ events for the $\gamma$-ray sources ($^{60}$Co, $^{137}$Cs, $^{22}$Na) and $^{252}$Cf neutron source, respectively, and $1\times{10}^6$ events for cosmogenic $^{12}$B, for each set of nonlinearity parameters.
Simulation statistics are over four times larger than the event statistics of the calibration data to enable energy scale model optimization while keeping computation time moderate. Identical event selection rules are applied to simulations, as previously described in section~\ref{sec:source}. MC spectra using nonlinear energy model parameter $k_{B}$ and $k_C$ from eq.~(\ref{eq:convert}) are subsequently rescaled by the overall scaling factor $A$. This allows the parameter $A$ to be varied without re-running the simulation. The $\chi^2$-based energy scale model optimization uses this simplification in the following section.

\subsection{\texorpdfstring{$\chi^2$-}{}based energy model optimization}\label{sec:chi2}
The energy response model of the P-I detector simulation is fit to the calibration data by minimizing the $\chi^2$ function~\cite{prd}:
\begin{equation}\label{eq:chi2}
    \chi^2_{\mathrm{data-MC}} = \sum_{\gamma} \chi^2_\gamma + \sum_{\mathrm{multi}}\chi^2_{\mathrm{multi}} + \chi^2_{^{12}\mathrm{B}},
\end{equation}
The $\chi^2$ function consists of contributions from both the energy spectra and the event multiplicity distributions from radioactive sources described in section~\ref{sec:source}. The $\chi^2$ contribution from the individual $\gamma$-ray spectra is denoted as $\chi^2_\gamma$. The event multiplicity distributions for each source are included in $\chi^2_{\mathrm{multi}}$. The intrinsic cosmogenic $^{12}$B $\beta$-spectrum contribution is denoted as $\chi^2_{^{12}\mathrm{B}}$. The $\gamma$-ray event multiplicity distribution captures the detector response to low energy deposits and threshold effects and thus plays an important role in constraining the energy model for highly-segmented inhomogeneous liquid scintillator detectors. The $^{12}$B $\beta$-dominated spectrum constrains the energy response model in the high energy region up to $\sim$14~MeV. 

\begin{figure}[ht]
\centering
\includegraphics[width=\textwidth]{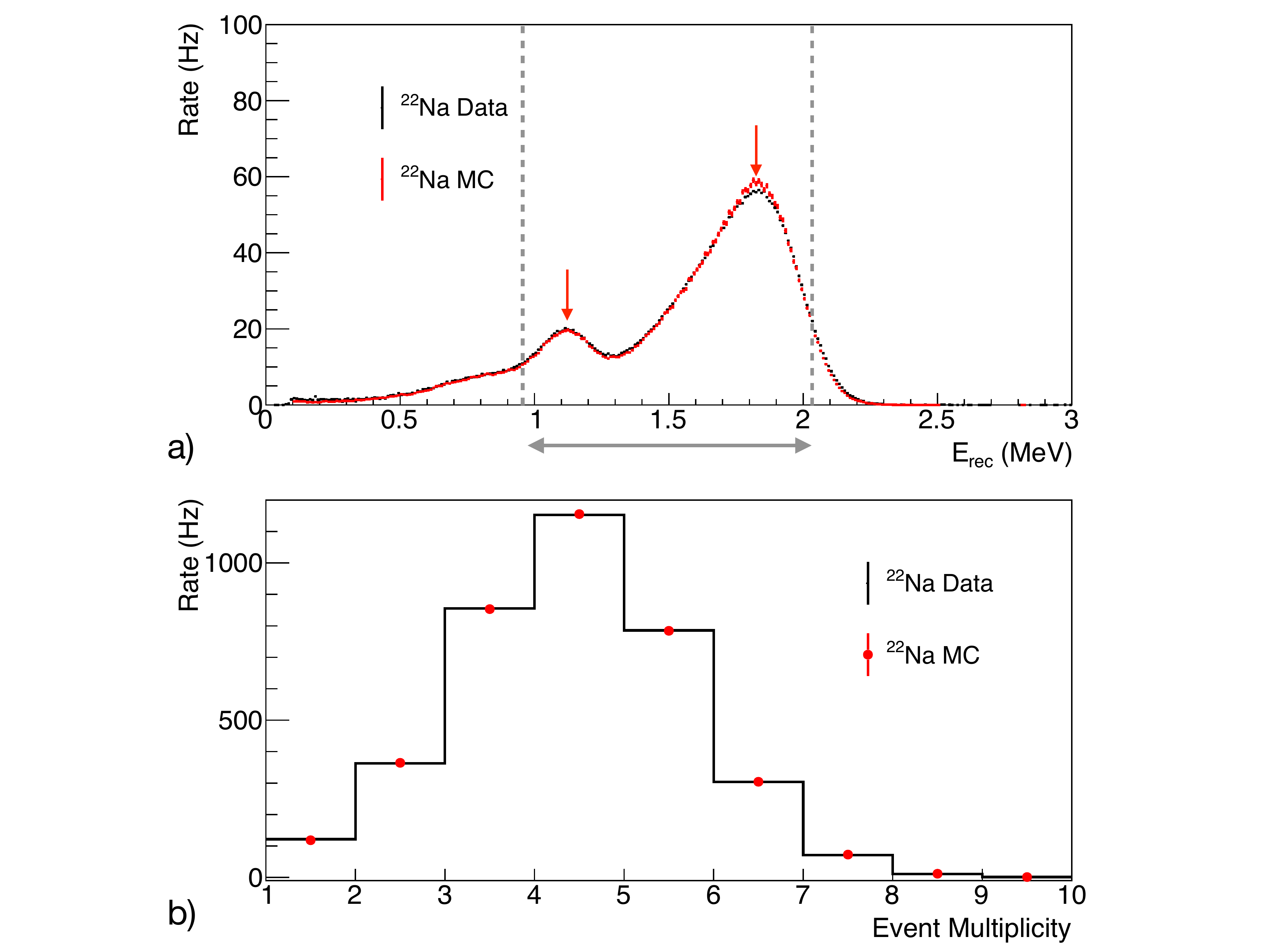}
\caption{ $^{22}$Na is one of several isotopes used for energy calibration. Here we show $^{22}$Na data and MC comparison for an internal energy scale model, corresponding to the best-fit set of nonlinearity model parameters. The reconstructed energy $E_\text{rec}$ represents scintillator quenched visible energy, which is shifted to lower energy compared to the true gamma energies. The gray lines in a) indicate the energy range for $\chi^2$ calculation. They are determined by the peak location and energy resolution as described above. The range for event multiplicity is between 1 and 9, as shown in b). }
\label{fig:spec_fit}
\end{figure}

A point-wise $\chi^2_{\mathrm{data-MC}}$ is calculated in a three-dimensional detector response parameter space ($A$,$k_{B}$,$k_C$) using weighted histogram comparison as follows~\cite{hist_chi2}:
\begin{equation}\label{eqn:hist_chi2}
    \chi^2 = \sum_{i=n_{\mathrm{min}}}^{n_{\mathrm{max}}}\frac{ (W_E w_{Si} - W_S w_{Ei} )^2 }{ W_E^2 \sigma_{Si}^2 + W_S^2 \sigma_{Ei}^2},
\end{equation}
where $W_E$ and $W_S$ are the integrated bin contents from experimental and simulation histograms, respectively, $w_i$ is the individual bin height, $\sigma_i$ is the individual uncertainty from each bin, $n_{\mathrm{min}}$ and $n_{\mathrm{max}}$ is the lowest and highest energy bin of the range of interest. The $\chi^2$ function is calculated in the energy range covering all peak areas and regions between the peaks. More specifically, all peaks in the energy spectrum from the data are identified by a peak-finding algorithm, and their lower and upper edges are located at $\pm3 \times$ energy resolution of 4.5 \%$/\sqrt{E}$ around each peak energy, which is the measured energy resolution from the P-I detector~\cite{prd}. The energy range of $\chi^2$ calculation for each type of source spectra is defined from the lower energy edge of the lowest energy peak to the upper edge of the highest energy peak. For example, figure~\ref{fig:spec_fit} shows $^{22}$Na calibration data in comparison to MC generated by an energy scale model corresponding to a set of nonlinearity model parameters (the best fit set is used here). For the energy distribution, the $\chi^2$ contribution from $^{22}$Na associated with this point is calculated in the energy range indicated by the gray lines in the top panel, with 0.01~MeV bin size. The event multiplicity range for $\chi^2_{\mathrm{multi}}$ calculation is between 1 and 9, beyond which event counts are negligible, as illustrated in the bottom panel of figure~\ref{fig:spec_fit}. 

\begin{figure}[ht]
\captionsetup[subfigure]{labelformat=empty}
\begin{subfigure}[b]{0.49\textwidth}
  \includegraphics[width=\textwidth]{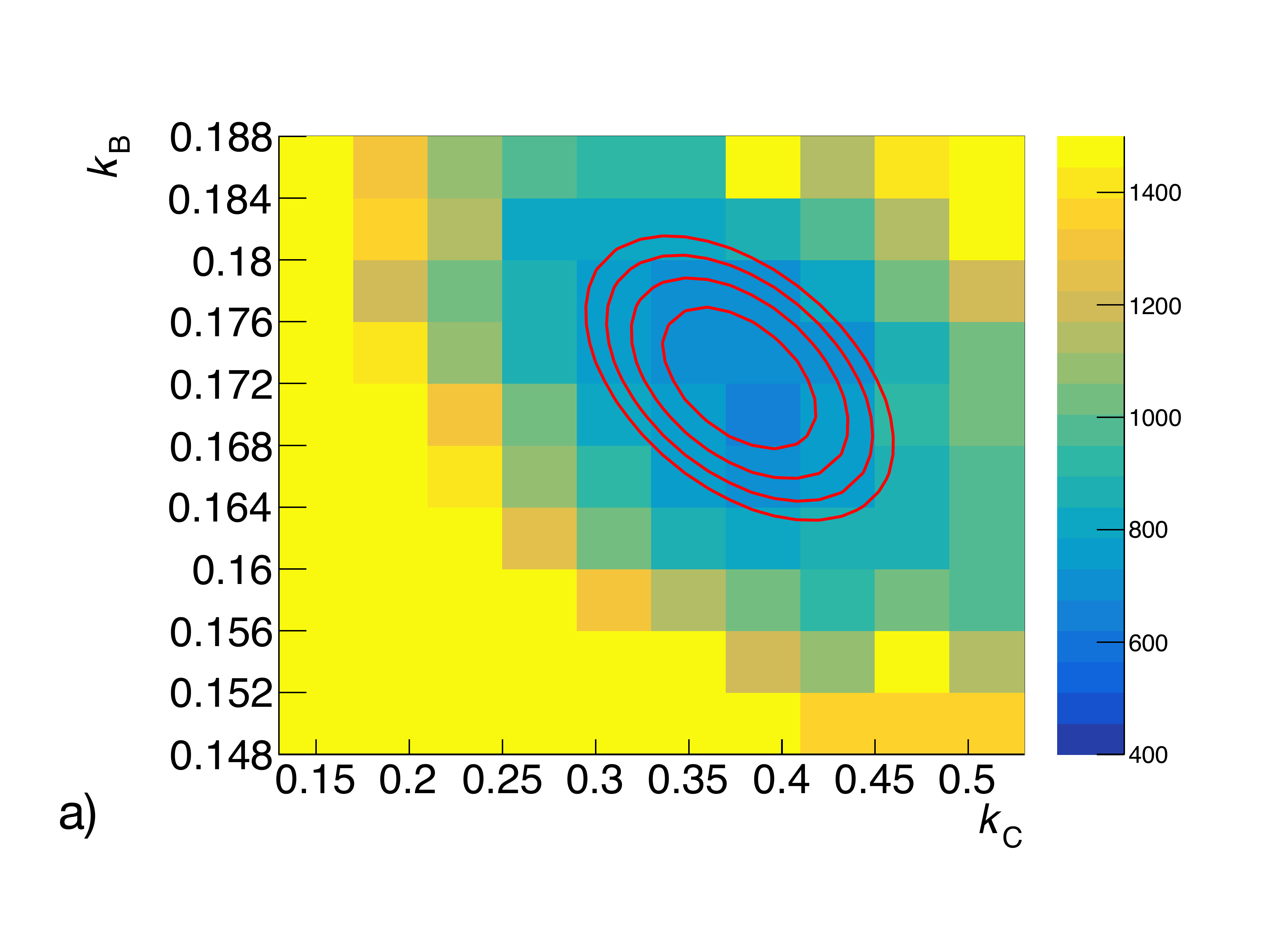}
\end{subfigure}
\hfill
\begin{subfigure}[b]{0.49\textwidth}
  \includegraphics[width=\textwidth]{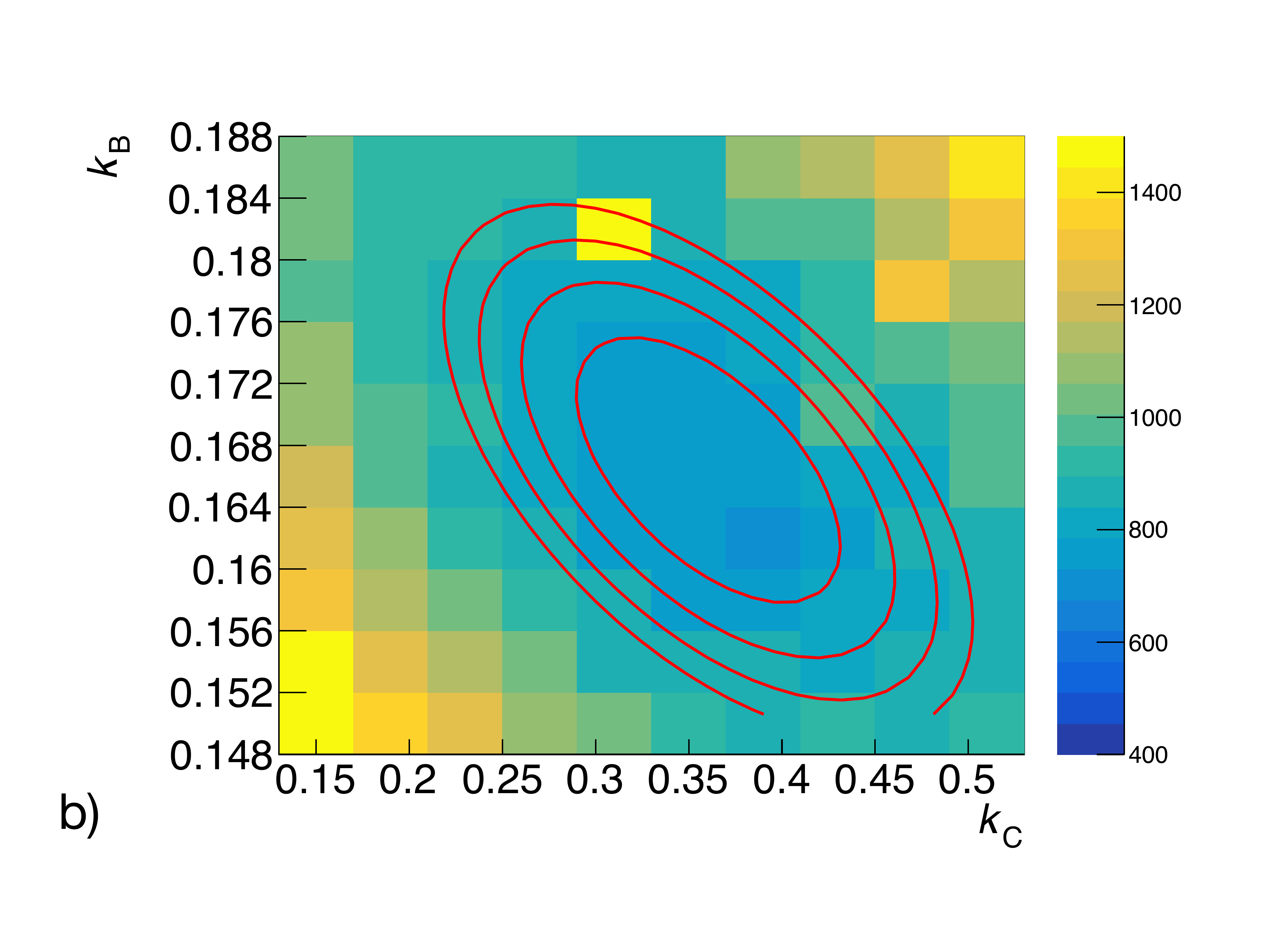}
\end{subfigure}
\caption{$\chi^2$ distribution in energy scale parameter space for a) internal calibration and b) external calibration. Each $\chi^2(k_B,k_C)$ on the grid represents the corresponding detector response deviation from the calibration data.  The contours are generated with $\Delta\chi^2 = 30$ increments for illustration purposes. Note the intervals in the contours are larger for the external calibration. Thus, the best-fit is less well-constrained compared to the internal case, but is still well-defined.}
\label{fig:chi2_map}
\end{figure}

The three-dimensional $\chi^2_{\mathrm{data-MC}}$ map in detector response parameter space ($A$,$k_{B}$,$k_C$) can be further simplified by optimizing the overall energy scaling factor $A$ once the other two parameters are set. This simplification does not change the fundamentals of the model optimization and significantly decreases the computation time for a point-wise simulation in axis-$A$. With $A$ decoupled from the three-dimensional response parameter space, a two-dimensional $\chi^2_{\mathrm{data-MC}}$ map can be easily visualized in the parameter coordinate ($k_{B}$,$k_C$) (figure~\ref{fig:chi2_map}). An elliptical contour with a $\Delta\chi^2 = 30$ relative to the minimum value is fitted to determine each parameter's best fit value and uncertainty. Best fit values of $k_{B}$ and $k_C$ parameters are the ellipse's center value, and their associated uncertainties are assigned as half of the projected width in each parameter axis. The best fit value of $A$ is the average of all values of $A$ within the contour, and the uncertainty is half of the $A$ variation. The choice of $\Delta\chi^2 = 30$ is intentionally conservative, chosen to be significantly larger than fluctuations due to the limited size of the simulation samples (approximately 10 units). The remaining  20 units corresponds to about 4$\sigma$ confidence intervals. Most importantly, the same $\Delta\chi^2 = 30$ is used for both internal and external energy scale model uncertainty analysis.

\section{Results and discussion} \label{sec:result}
Using the data and methods detailed in the previous section, here we summarize the result of a P-II type external calibration within the context of existing P-I analysis. The possible inclusion of an AmBe neutron source in P-II calibration is also investigated. Additionally, the estimated systematic uncertainty from the external energy scale model is compared to projected P-II statistical uncertainty. 
\subsection{Comparison of Internal and External Calibration Energy Scale Uncertainties}\label{res:main}
Following the prescription described in section~\ref{sec:chi2}, table~\ref{table:nonlin} summarizes the best fit values of each parameter and associated uncertainties. While the external calibration energy model is compatible with the internal calibration for all parameters within the assigned uncertainty, it is less well constrained by approximately a factor of two. We will discuss the implications of this model parameter uncertainty in terms of systematic contribution to spectrum uncertainty in section~\ref{sec:sys} and how to reduce parameter uncertainties with additional high-energy $\gamma$-rays generated from an AmBe neutron source in section~\ref{sec:AmBe}. Given the reduced statistics of calibration events in the external configuration, the results presented here are conservative and improvements, such as higher activity sources and longer calibration acquisition time, are expected from an actual P-II detector setup.

\begin{table}[ht]
\centering
\caption{Nonlinear energy model fit parameters. The number of degrees of freedom (ndf) is different in two configurations due to change of region of interest in the energy spectrum. }
\setlength{\tabcolsep}{2pt} % default value is 6pt
\begin{tabular}{c c c c c c}
\hline\hline
Configuration & $A$ & $k_B$ (cm/MeV) & $k_C$ & $\chi^2$ & ndf \\  \hline
Internal & 1.0080 $\pm$ 0.0020  & 0.1723 $\pm$ 0.0038 & 0.3772 $\pm$ 0.0338 & 657.6 & 330\\ 
External & 1.0073 $\pm$ 0.0055 & 0.1664 $\pm$ 0.0070 & 0.3603 $\pm$ 0.0579 & 712.4 & 460\\ 
\hline
\end{tabular}

\label{table:nonlin}
\end{table}

\begin{figure}[h]
\captionsetup[subfigure]{labelformat=empty}
\begin{subfigure}[b]{0.49\textwidth}
   \includegraphics[width=\textwidth]{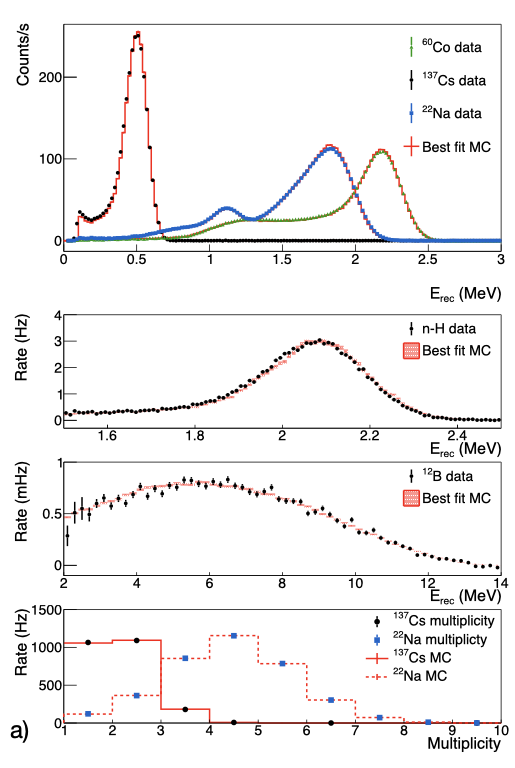}
\end{subfigure}
\hfill
\begin{subfigure}[b]{0.49\textwidth}
   \includegraphics[width=\textwidth]{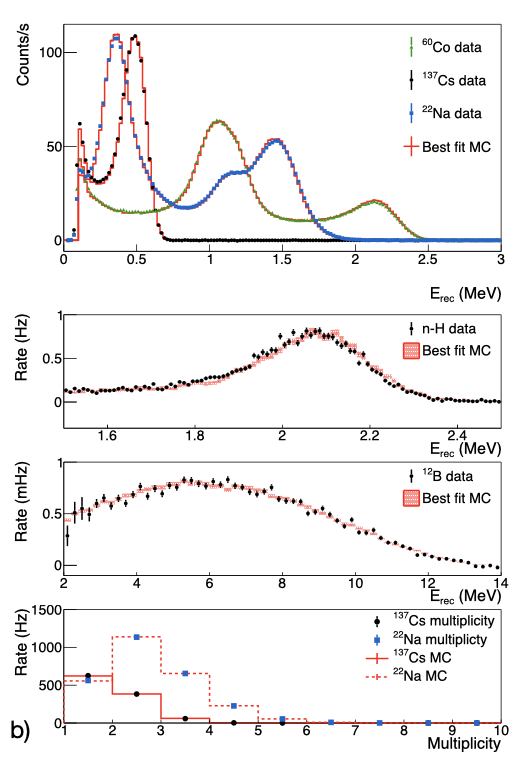}
\end{subfigure}
\caption{Data-MC comparison of calibration source spectra and event multiplicity for a) best-fit internal energy scale model and b) best-fit external energy scale model. The reconstructed energy $E_\text{rec}$ represents scintillator quenched visible energy, which is shifted to lower energy compared to the true gamma energies. External calibration events generally have lower event statistics and decreased event multiplicity due to smaller solid angle coverage. The external calibration spectra show more pronounced single-$\gamma$-ray photopeaks, with lower rates of total $\gamma$-ray photopeaks. The total $\chi^2$ from both energy scale models are listed in table~\ref{table:nonlin}.}
\label{fig:spe_result}
\end{figure}

Using the best fit values in table~\ref{table:nonlin} for internal and external calibration, both models demonstrate good agreement between data and MC in the reconstructed energy spectrum and event multiplicity, as shown in figure~\ref{fig:spe_result}. Noticeable changes in both the spectral shape and multiplicity are apparent between internal and external configurations. The detector active volume only covers half of the solid angle of the external sources and thus is less likely to fully capture multiple $\gamma$-rays, causing the change of energy spectrum and downshift in event multiplicity shown in figure~\ref{fig:spe_result}. nH capture $\gamma$-ray rates are reduced due to less neutron captures internally, similar to the $\gamma$-ray source. The event multiplicity of nH event will suffer a similar downshift, depending on how deep the nH reaction happens in the fiducial volume.
On the other hand, $^{12}$B, generated intrinsically via the $^{12}$C(n,p)$^{12}$B reaction in the liquid scintillator, is independent of the calibration setup. In the external calibration setup, both the $\gamma$-ray calibration sources and the 2.2~MeV $\gamma$-ray from neutron captures on hydrogen see a reduction of event statistics that could be compensated by higher activity sources in the actual P-II calibration.
\subsection{Inclusion of AmBe source}\label{sec:AmBe}
The PROSPECT collaboration acquired an AmBe source near the end of the P-I data-taking period. An AmBe source is of particular relevance to IBD neutrino detectors. Neutrons and a correlated 4.4~MeV $\gamma$-ray are produced in the AmBe source via the $^{9}$Be($\alpha$,n)$^{12}$C* reaction. The correlated pair of neutron and $\gamma$-ray is similar to IBD events and can be used as a tool to quantify neutron capture efficiency and neutron mobility. They also provide a 4.4~MeV energy calibration feature, further constraining $k_B$ and $k_C$ over the high-energy region.

\begin{figure}[h]
\centering
\includegraphics[width=\textwidth]{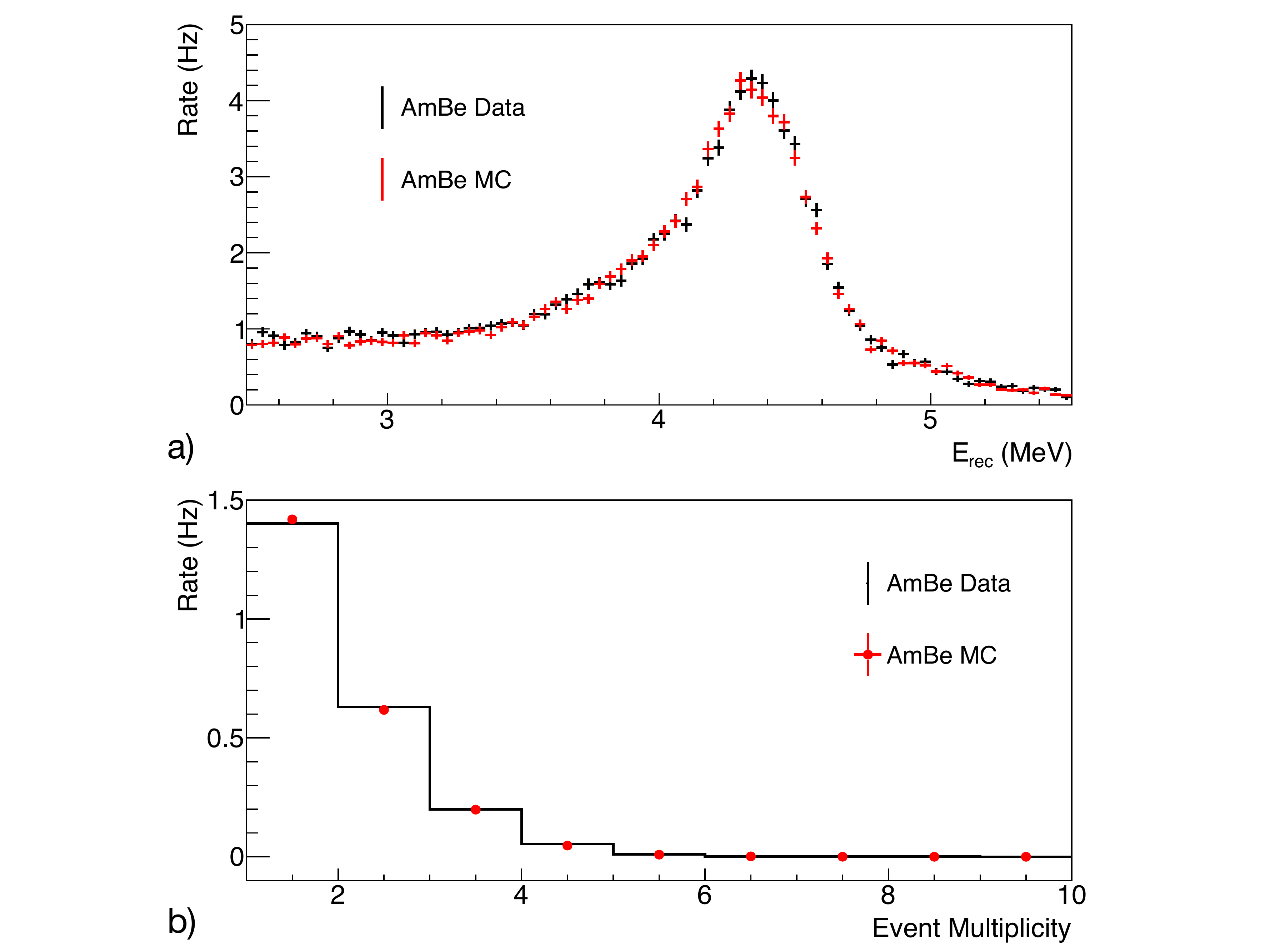}
\caption{AmBe 4.4~MeV $\gamma$-ray energy spectrum in a) and event multiplicity distribution in b) using the external energy scale model obtained with AmBe calibration.}
\label{fig:AmBe}
\end{figure}

The AmBe source was not included in the initial calibrations and was only used as a cross-check for energy scale model data-MC comparison in previously published results~\cite{prd}. However, we include these data here to test the impact of higher energy calibration features on future calibration campaigns. Inclusion of the AmBe source in the calibration data analysis is accomplished via the equation,
\begin{equation}\label{eq:ambe}
    \chi^2_{\mathrm{tot}} = \chi^2_{\mathrm{data-MC}} + [ \chi^2_{\gamma} + \chi^2_{\mathrm{multi}} ]_{\mathrm{AmBe}},
\end{equation}
where $\chi^2_{\mathrm{data-MC}}$ is defined in eq.~(\ref{eq:chi2}) and the $\chi^2$ contributions from the AmBe 4.4~MeV $\gamma$-ray reconstructed energy spectrum and event multiplicity are added to $\chi^2_{\mathrm{tot}}$. The best-fit energy scale model with the AmBe source shows good agreement between the AmBe data and MC as shown in figure~\ref{fig:AmBe}, with $\chi^2_{\mathrm{AmBe}}$/ndf = 250.6/245. The fit for other sources shows good agreement as well. For example, $^{22}$Na energy spectrum contributes $\chi^2_{\mathrm{Na}}$/ndf = 254.7/106 with the inclusion of the AmBe source, compared to $\chi^2_{\mathrm{Na}}$/ndf = 254.3/106 without the AmBe source. The benefit of including an energetic calibration $\gamma$-ray is an improved constraint on the higher energy calibration. In particular, the $k_C$ uncertainty is reduced, as shown in table~\ref{table:AmBe}, where we see a 31 \% improvement. The amount of the improvement is unlikely to increase in P-II calibration run, given that the AmBe data we used contains sufficient events statistics for the purpose of the study.
\begin{table}[ht]
\centering
\caption{Reduced $k_C$ uncertainties with the inclusion of AmBe source in the external calibration.}
\setlength{\tabcolsep}{5pt} % default value is 6pt
\begin{tabular}{c c c c }
\hline\hline
Configuration & $A$ & $k_B$ (cm/MeV)  & $k_C$  \\ [0.5ex] \hline
Internal & 1.0080 $\pm$ 0.0020  & 0.1723 $\pm$ 0.0038 & 0.3772 $\pm$ 0.0338 \\ 
External & 1.0073 $\pm$ 0.0055 & 0.1664 $\pm$ 0.0070 & 0.3603 $\pm$ 0.0579 \\ 
External with AmBe & 1.0028 $\pm$ 0.0045  & 0.1658 $\pm$ 0.0061 & 0.4055 $\pm$ 0.0399 \\[0.5ex]
\hline
\end{tabular}
\label{table:AmBe}
\end{table}

\subsection{Systematic uncertainty}\label{sec:sys}
A key performance metric of the calibration approach is the contribution to the overall systematic uncertainty from the energy scale model. We desire that the systematic uncertainty contribution from the external calibration scheme will be less than or equal to the anticipated statistical uncertainties of P-II. To evaluate the impact of energy scale systematic uncertainty on the oscillation and neutrino spectrum physics goals of P-II, we begin by calculating a covariance matrix. 

The covariance matrix is generated by comparing a set of toy model spectra to a reference IBD spectrum. Reference IBD spectra are generated by passing the Huber-Mueller model~\cite{Huber,Mueller} predicted neutrino spectrum through the detector response with the best-fit value of the parameters, and toy model spectra use Gaussian distributed parameters with sigma values assigned by the corresponding error bar. Toy models are generated with uncorrelated uncertainties. Reference IBD spectra consist of $4\times{10}^7$ simulated IBD interactions. Toy model spectra consist of $1\times{10}^3$ toy simulations, each of $4\times{10}^6$ simulated IBDs. Approximately 10 \% of simulated IBDs pass the selection cuts in the fiducial volume.

\begin{figure}[h]
\centering
\includegraphics[width=0.8\textwidth]{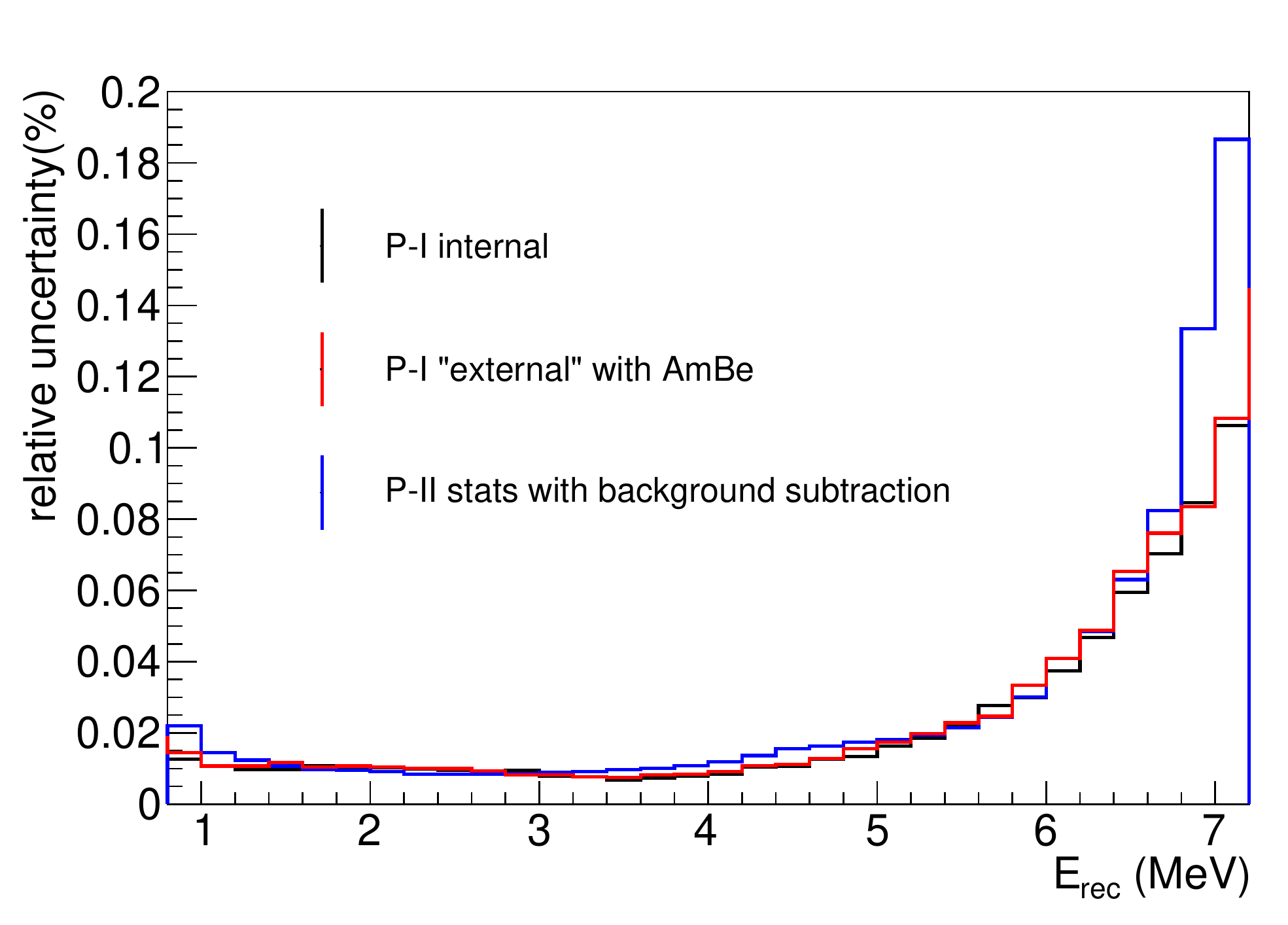}
\caption{Relative systematic uncertainty from the energy scale model is plotted in the energy range where most IBD events fall. P-I internal in black represents the uncertainty from the P-I internal calibration without the AmBe source. P-I "external" in red represents the uncertainty from the P-II type external calibration study with the AmBe source. P-II signal statistics with background subtraction in blue includes the estimated statistical uncertainty from the expected IBD statistics in the P-II detector and estimated background statistics. }
\label{fig:CovMat}
\end{figure}

The diagonal elements of the reduced uncertainty covariance matrix provides a measure of the relative uncertainty in each energy bin. In figure~\ref{fig:CovMat}, the relative uncertainty is plotted between 0.8~MeV and 7.2~MeV visible energy where most reactor antineutrino events from the IBD process fall. The external calibration uncertainty with the AmBe source (red line) shows a similar level of performance to the P-I internal calibration without the AmBe source (black line). P-II is expected to observe more than $3.5\times{10}^5$ IBD events during two years' deployment, with an estimated signal-to-background ratio of 4.3. Estimated P-II statistical uncertainty with background subtraction is plotted in blue, slightly above the uncertainty of the external calibration. Further reduction of the external calibration systematic uncertainties can be achieved with higher activity sources and longer calibration acquisition time, resulting in smaller systematic uncertainties than the estimated two years' P-II neutrino statistics with background subtraction. This study suggests that the energy scale model systematic uncertainties will not be dominant for the improved P-II detector~\cite{p2physics}. 

\subsection{Additional Studies}
We have demonstrated an external calibration scheme viable for the upgraded P-II detector. Here we continue to briefly address some specific questions related to the design of P-II: the detector-source stand-off distance, increased backgrounds in the location radioactive sources will be placed, and the impact of inter-segment cross-talk. The detector's size and geometry, and the dimensions of each segment can also impact the performance of the external calibration technique. However, we do not expect significant impact on calibration due to the small change in the segment dimensions.

\begin{figure}[h]
\centering
\includegraphics[width=\textwidth]{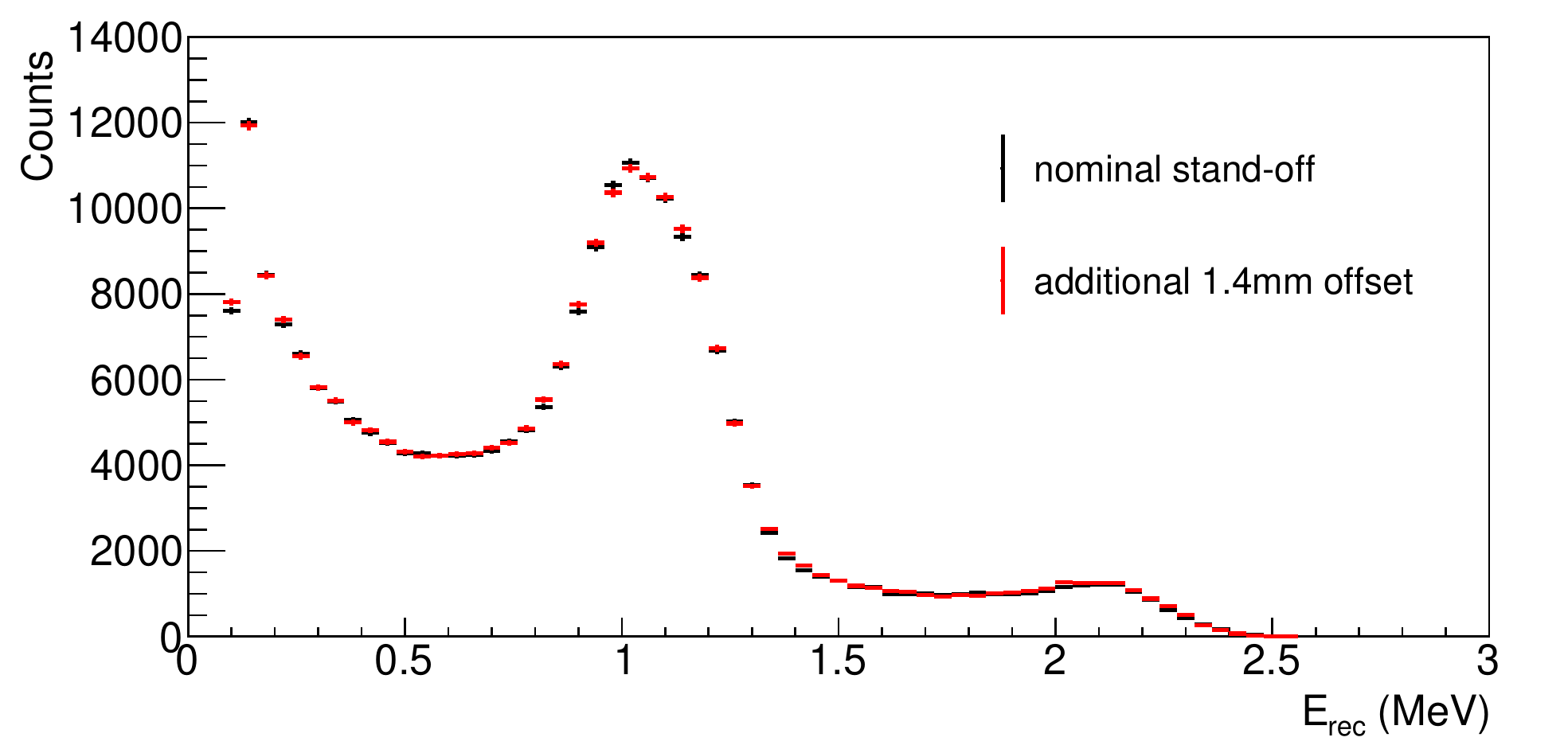}
\caption{ Simulated $^{60}$Co spectrum externally deployed with a stand-off distance of 1.4~mm further away from the nominal position.  }
\label{fig:standoff}
\end{figure}

Having radioactive sources deployed externally introduces a stand-off distance between the source and the boundary of the active liquid scintillator volume, which can be treated as an additional fitting parameter if the transverse source location is not precisely determined. The preliminary P-II design introduces a stand-off distance of approximately 3~cm. We evaluate the precision requirement of stand-off distance by comparing MC spectra at different stand-off distances to the nominal position spectrum. In figure~\ref{fig:standoff}, MC simulations of additional source stand-off distances of 1.4~mm further away from the nominal position show negligible differences ($\chi^2/$ndf= 250.6/245) in the spectrum. We further increase the offset to 2, 5 and 10 times the unit of 1.4~mm, which corresponds to 1\% of the segment width. At those offset positions, apparent spectral deviations are visible with increasing $\chi^2/$ndf values, 1.25, 1.74 and 4.19 respectively. A positioning requirement of $\pm$1.4~mm or $\pm$1\% of the segment width is achievable through careful engineering and will have minimal impact on the desired energy scale uncertainty. 

By definition, an external calibration scheme will be primarily using data from segments on edge, and these are likely to experience higher $\gamma$-ray background rates than those internal to the detector. In the P-II detector, an externally deployed source sits outside the active liquid scintillator but inside P-II exterior shielding packages. The P-I detector observed increased $\gamma$-ray background rates by up to a factor of four in the outer layers of the detector segments compared to the detector interior. It is these outer layers that are primarily calibrated by external sources. By artificially increasing the background rates by a factor of four, we find these increased $\gamma$-ray backgrounds translate to increased uncertainties in the reconstructed energy spectrum and event multiplicity by $\sim$3-4 \%. These increased uncertainties are negligible in the $\chi^2$-optimization process and could be mitigated by using a higher activity source or better P-II exterior shielding package.

The P-II detector separates the PMTs from the active liquid scintillator volume by an acrylic window. This design change introduces $\sim$0.5 \% cross-talk between adjacent segments and $\sim$0.01 \% cross-talk between diagonal segments, estimated with optical MC simulations and benchtop testing. To investigate the impact of such cross-talk on calibration performance, various levels of artificial cross-talk are added to the P-I calibration data and simulation, and the energy scale analysis is repeated. The artificial cross-talk takes into account the non-uniform longitudinal dependence along the segment based on the optical simulations. The magnitude of the cross-talk increases non-linearly by about 80 \% as the light source moves from the center to the end of a segment. Using an internal calibration setup (similar results for external calibration setup), the effective energy scale model with 2 \% cross-talk shows no significant difference in the energy response parameters, as shown in table~\ref{table:xtalk}. 

A specific concern is that the level of cross-talk will not be exactly known from simulations and {\it in-situ} measurements. To investigate the precision required, different levels of cross-talk are applied to the data and MC. We obtain a similar effective energy scale model using an artificial cross-talk of 2 \% applied to data and 2.5 \% applied to simulation. In this case, a small increase in the $A$ parameter is observed, to compensate for more light lost to neighboring segments (where it falls below the per-segment detection threshold) in the simulation than the data. This indicates a tolerance of up to 0.5 \% for cross-talk measurement precision. The energy scale models with various levels of cross-talk are summarized in table~\ref{table:xtalk}; they all show good agreement in energy spectrum and event multiplicity.

\begin{table}[h]
\centering
\caption{Effective energy scale model with various level of cross-talk in an internal calibration setup. External calibration with cross-talk yield similar results.}
\setlength{\tabcolsep}{4pt} % default value is 6pt
\begin{tabular}{c c c c }
\hline\hline
Configuration & $A$ & $k_B$ (cm/MeV)  & $k_C$  \\ [0.5ex] \hline
No cross-talk & 1.0080 $\pm$ 0.0020  & 0.1723 $\pm$ 0.0038 & 0.3772 $\pm$ 0.0338 \\ 
2 \% sim \& 2 \% data & 1.0100 $\pm$ 0.0020  & 0.1734 $\pm$ 0.0037 & 0.3783 $\pm$ 0.0357 \\[0.5ex]
2.5 \% sim \& 2 \% data & 1.0130 $\pm$ 0.0010  & 0.1744 $\pm$ 0.0036 & 0.3720 $\pm$ 0.0350 \\[0.5ex]
\hline
\end{tabular}
\label{table:xtalk}
\end{table}

\section{Conclusion}
\label{sec:conclusion}
P-II, an upgrade of the P-I detector, introduces design improvements that aim to provide long-term stable operation and increase the target mass within a fixed footprint by more efficient use of the internal volume. As a result, previous internal calibration access is no longer available. We have demonstrated, by using P-I data and a well-benchmarked MC package, that an external source calibration strategy for P-II is feasible. A P-II type external calibration study shows a performance degradation in constraining parameters by a factor of less than two. The inclusion of an AmBe neutron source, which was previously only used for detector energy response validation in P-I, brings the performance of the external calibration on par with the internal calibration without an AmBe source. The estimated systematic uncertainty from the external energy scale model will meet the physics goals of the program. Additional studies on topics specific to P-II provide design guidance on items like detector-source stand-off distance, increased $\gamma$-ray backgrounds at external calibration locations, and inter-segment crosstalk. These findings may have value for others designing segmented reactor antineutrino detectors for basic science or applications.

\acknowledgments
This material is based upon work supported by the following sources: US Department of Energy (DOE) Office of Science, Office of High Energy Physics under Award No. DE-SC0016357 and DE-SC0017660 to Yale University, under Award No. DE-SC0017815 to Drexel University, under Award No. DE-SC0008347 to Illinois Institute of Technology, under Award No. DE-SC0016060 to Temple University, under Award No. DE-SC0010504 to University of Hawaii, under Contract No. DE-SC0012704 to Brookhaven National Laboratory, and under Work Proposal Number SCW1504 to Lawrence Livermore National Laboratory. This work was performed under the auspices of the U.S. Department of Energy by Lawrence Livermore National Laboratory under Contract DE-AC52-07NA27344 and by Oak Ridge National Laboratory under Contract DE-AC05-00OR22725. Additional funding for the experiment was provided by the Heising-Simons Foundation under Award No. 2016-117 to Yale University.

J.G. is supported through the NSF Graduate Research Fellowship Program. This work was also supported by the Canada First Research Excellence Fund (CFREF), and the Natural Sciences and Engineering Research Council of Canada (NSERC) Discovery  program under grant RGPIN-418579, and Province of Ontario.

We further acknowledge support from Yale University, the Illinois Institute of Technology, Temple University, University of Hawaii, Brookhaven National Laboratory, the Lawrence Livermore National Laboratory LDRD program, the National Institute of Standards and Technology, and Oak Ridge National Laboratory. We gratefully acknowledge the support and hospitality of the High Flux Isotope Reactor and Oak Ridge National Laboratory, managed by UT-Battelle for the U.S. Department of Energy.

\bibliographystyle{JHEP}
\bibliography{main}

\end{document}